\documentclass[aps,prb,twocolumn,superscriptaddress,showpacs]{revtex4}

\usepackage{graphicx}

\begin{document}

\title{Wave-packet view of the Dirac electron }

\author{Chih-Piao Chuu}
\affiliation{Department of Physics, University of Texas at Austin,
Austin, TX 78712}
\author{Ming-Che Chang}
\affiliation{Department of Physics, University of Texas at Austin,
Austin, TX 78712}
\affiliation{Department of Physics, National
Taiwan Normal University, Taipei, Taiwan 11677}
\author{Qian Niu}
\affiliation{Department of Physics, University of Texas at Austin,
Austin, TX 78712}
\date{\today}

\begin{abstract}
By viewing the electron as a wavepacket in the positive energy
spectrum of the Dirac equation, we are able to achieve a much
clearer understanding of its behavior under weak electromagnetic
fields.  The intrinsic spin magnetic moment is found to be
established from the self-rotation of the wavepacket. A
non-canonical structure is also exhibited in the equations of
motion due to non-Abelian geometric phases of the Dirac spinors.
The wavepacket energy can be expressed simply in terms of the
kinetic, electrostatic, and Zeeman terms only. This can be
transformed into an effective quantum Hamiltonian by a novel
scheme, and reproduces the Pauli Hamiltonian with all-order
relativistic corrections.
\end{abstract}

 \maketitle

When the electron spin was first discovered, Uhlenbeck and
Goudsmit regarded that it came from the self-rotation of the
electron charge sphere\cite{uhlenbeck}. However, this idea was
soon swept off by Lorentz, who argued that the surface of sphere
would rotate with a tangential speed at 137 times of the speed of
light in order to produce the spin angular moment of
$\hbar/2$\cite{pais}.  One is thus forced to accept the spin as a
discrete degree of freedom in abstract form, and eventually finds
comfort in the Dirac theory\cite{dirac}, which explains the atomic
spectra with great accuracy.

In the non-relativistic limit, the electron spin manifests through
its physical attribute, an intrinsic magnetic moment, called the
Bohr magneton.  It is responsible for the Zeeman splitting of
atomic spectral lines in a magnetic field.   The spin also couples
with orbital motion in the presence of an electric field, such as
the Coulomb field inside an atom, which is responsible for the
fine structures of the atomic spectra.  Such effects are nicely
captured in the Pauli Hamiltonian, which can be rigorously derived
from the Dirac equation through the Foldy-Wouthuysen (FW)
transformation\cite{foldy}.  However, the appearance of the
intrinsic magnetic moment for the spin has been mysterious, and
the usual physical picture linking the spin-orbit energy to the
Zeeman effect has been rather cumbersome:  A simple
reference-frame argument misses a factor of two, and Thomas
precession has to be taken into account to make the story
complete\cite{thomas}.

In this paper, we take an alternative point of view to regard the
electron as a wavepacket in the positive energy spectrum of the
Dirac equation. The wavepacket has a minimum radius of the Compton
wavelength, which is larger by a factor of 137 than the so-called
classical radius of the electron, the size that Lorentz used in
his argument. In our viewpoint, the wavepacket is found to be
self-rotating, with an orbital angular momentum twice of the spin,
and produces a magnetic moment of the Bohr magneton with its
circulating charge. Therefore, the heuristic picture of Uhlenbeck
and Goudsmit is essentially correct, and the so-called intrinsic
magnetic moment of the spin originated from the orbital motion at
the fundamental level.

In the presence of weak electromagnetic fields, it is possible to
formulate a semiclassical dynamics for the variables of the center
of charge, kinetic momentum and the average
spin\cite{bliokh,teufel,other}. The magnetic moment gives the
Zeeman term as a correction to the wavepacket energy, but the
spin-orbit energy is notably absent. The spin-orbit term can arise
only because of a shift from the physical position variable to an
effective but unphysical variable in the Pauli Hamiltonian
\cite{blount,yafet}, which is closely related to the non-abelian
geometric phase of the Dirac spinors\cite{mathur}.  In this work,
we formulate a generalization of the Peierls
substitution\cite{peierls}, which transforms the wavepacket energy
to the Pauli Hamiltonian for arbitrary momenta\cite{silenko} by a
novel quantization procedure. In our treatment the spin-orbit
coupling is a natural product in the canonicalization of the
physical variables, and its form is found to be gauge dependent.
We also suggest that, for a particle with anomalous magnetic
moment, its mean position calculated from the expectation value of
the position operator is actually not the true center of charge,
and a correction term is required in predicting the correct
trajectory of the particle.

We now proceed with the calculations leading to the above results.
We first construct a wavepacket from the positive part of the
energy spectrum\cite{culcer,shindou},
\begin{equation}
|w\rangle=\int d^3 \textbf{q}
a(\textbf{q},t)e^{i\textbf{q}\cdot\textbf{r}}
[\eta_1(\textbf{q},t)|u_1\rangle+\eta_2(\textbf{q},t)|u_2\rangle],\label{packet}
\end{equation}
where $\eta_i$ are the probability amplitudes of finding the
electron in the two bands, and the Dirac electron wavepacket
constitutes of linearly combination of spin up and spin down
amplitudes of a free particle with a localized distribution
function $a(\textbf{q},t)$ centered at $\textbf{q}_c$ in momentum
space. The phase of $a(\textbf{q},t)$ is chosen to ensure that the
mean position of the wavepacket is located at $\textbf{r}_c$.  The
conventional Dirac spinors $|u_i\rangle$ are obtained from a gauge
rotation $e^{\beta\mbox{\boldmath$\alpha$}\cdot{\hat q}\omega/2}$
from $(1000)^{T}$ and $(0100)^{T}$, in which
$\mbox{\boldmath$\alpha$}$ and $\beta$ are the familiar Dirac
matrices, and $\tan\omega=\frac{\hbar q}{mc}$.\cite{bjorken}

The intrinsic minimum size of electron wavepacket can be obtained
by introducing the projection operators, $P$ and $Q$, for the
electron and positron parts of the energy spectrum respectively,
with $P+Q=1$. The mean square radius of the wavepacket can then be
written as\cite{marzari}
\begin{equation}
(\Delta r)^2=\langle \textbf{r}P\textbf{r}\rangle-\langle
\textbf{r}\rangle^2+\langle \textbf{r}Q\textbf{r}\rangle,
\end{equation}
the wavefunction $|w\rangle$ is an eigenstate of $P$ with
eigenvalue 1. Using this fact, one can see that $\langle
\textbf{r}P\textbf{r}\rangle-\langle \textbf{r}\rangle^2$ is
simply the mean squared radius for the projected position operator
$P\textbf{r}P$, and therefore is positive definite. The remaining
part
\begin{equation}
\langle\textbf{r}Q\textbf{r}\rangle=\int
d^3q|a(\textbf{q})|^2\left[\frac{\lambda_c}{2\epsilon^2(q)}\right]^2
\end{equation}
is also positive definite, where the Compton wavelength
$\lambda_c=\hbar/mc$ is involved and
$\epsilon({q})\equiv\sqrt{1+(\hbar q/mc)^2}$. If we consider a
wavepacket with a relatively sharp shape in momentum space, then $
\langle\textbf{r}Q\textbf{r}\rangle^{1/2}=\frac{\lambda_c}{2\epsilon^2({
q}_c)}$, which reduces to $\lambda_c/2$ as $q_c \to 0$. This may
be regarded as the intrinsic size of the non-relativistic
electron. If one wishes to probe length scales smaller than the
Compton wavelength, the positron spectrum must be involved in the
wavepacket and therefore the physical entity being probed is no
longer a pure electron.

Within this framework, the electron wavepacket is found to be
rotating about its center of charge in general, and this rotation
is responsible for the intrinsic magnetic moment usually
associated with the spin.
\begin{figure}
\center
\includegraphics[width=4in]{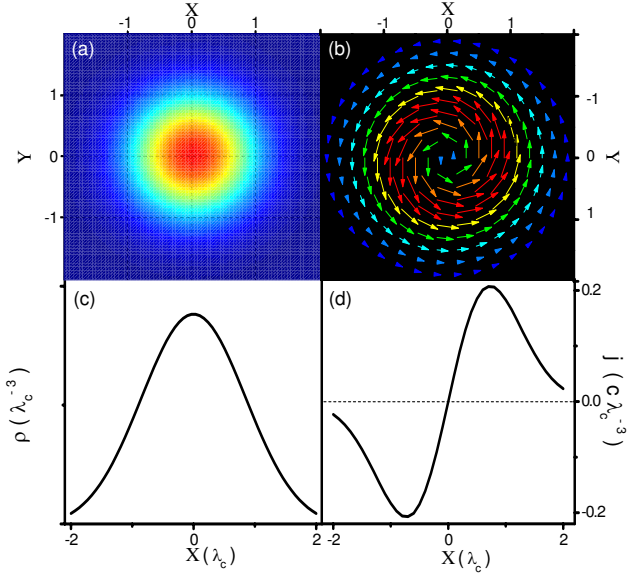}
\caption{(Color online) Distributions of (a) particle density
$\rho=w^\dagger(\textbf{r})w(\textbf{r})$ and (b) particle current
density
$\textbf{j}(\textbf{r})=w^\dagger(\textbf{r})c\mbox{\boldmath$\alpha$}w(\textbf{r})$
 on the $X$-$Y$ plane for an electron wavepacket, the spin is oriented out of the
plane. The units are in $\lambda_c$. The colors indicate the
magnitude of the density from low (blue) to high (red). The (c)
and (d) are profiles of (a) and (b) along the $X$-axis.}
\label{ChuuFig1}
\end{figure}
The Fig.1 shows the probability distribution and the current
density of an electron wavepacket with a spin oriented out of the
plane in the frame of wavepacket. An orbital circulation around
the spin axis can be clearly seen from this figure. The mechanical
angular momentum for the orbital circulation can be calculated
from the formula
$\textbf{L}=m\langle(\textbf{r}-\textbf{r}_c)\times
\textbf{v}\rangle =m\int d^3r (\textbf{r}-\textbf{r}_c)\times
\textbf{j}(\textbf{r})$. A general analytic formula for this
quantity has been derived in Ref.\cite{culcer}, which has the
following form $\textbf{L}=\eta^\dagger\mbox{\boldmath${\cal
L}$}\eta$, where $\eta=(\eta_1,\eta_2)^T$ is a two-component
spinor, and
\begin{equation}
\mbox{\boldmath${\cal L}$}_{ij}(\textbf{q}_c)=\sum_{l=3,4}m\langle
u_i|i\frac{\partial}{\partial{\textbf
q}_c}|u_l\rangle\times\langle u_l|
c\mbox{\boldmath$\alpha$}|u_j\rangle\label{L}.
\end{equation}
Notice that indices $i$ and $j$ refer to positive energy levels,
where index $l$ runs over the negative energy levels. A
straightforward calculation of Eq.~(\ref{L}) yields
\begin{equation}
\mbox{\boldmath${\cal
L}$}=\frac{\hbar}{\epsilon^2}\left(\mbox{\boldmath$\sigma$}+\lambda_c^2\frac{{\textbf
q_c \cdot{\mbox{\boldmath$\sigma$}}}}{\epsilon+1}{\textbf
q}_c\right),
\end{equation}
where $\mbox{\boldmath$\sigma$}$ are the Pauli matrices. It can be
shown that a natural connection exists between the self-rotating
angular momentum and the projected spin,
\begin{eqnarray}
\hbar\mbox{\boldmath$\tau$}_{ij}(\textbf{q}_c)\equiv\hbar\langle
u_i|\mbox{\boldmath$\Sigma$}|u_j\rangle=\epsilon\mbox{\boldmath${\cal
L}$}_{ij}(\textbf{q}_c),
\end{eqnarray}
where $\mbox{\boldmath$\Sigma$}$ are $4\times 4$ spin matrices. At
extremely high velocity($\epsilon\rightarrow
\lambda_c\textbf{q}_c$), $\mbox{\boldmath${\cal L}$}$ reduces to
zero but $\mbox{\boldmath$\tau$}$ approaches the projection along
the direction of momentum $({\hat q
}_c\cdot\mbox{\boldmath$\sigma$}){\hat q}_c$.

The magnetic moment of the wavepacket is similarly defined as
$\textbf{M}=-{e\over 2}\langle (\textbf{r}-\textbf{r}_c)\times
\textbf{v}\rangle =-{e\over 2}\int d^3r
(\textbf{r}-\textbf{r}_c)\times \textbf{j}(\textbf{r})$. This
differs from the mechanical angular momentum of self-rotation by
the factor of $e/2$ in place of $m$, which ensures the usual
relation $\textbf{M}=-(e/2m)\textbf{L}$ to be hold. One thus has
\begin{equation}
{\textbf M}=-\frac{ge}{2\epsilon
m}\frac{\hbar\mbox{\boldmath$\tau$}}{2},
\end{equation}
where $g=2$. It is very interesting to see that the so-called
intrinsic magnetic moment of the spin really comes from the
self-rotation of the wavepacket. In this sense, Uhlenbeck and
Goudsmit's view of the electron spin as a rotating sphere is
meaningful. The sphere is rigid near the core in the sense that
the rotation velocity of the wavepacket is approximately
proportional to the radius near the center.

The spin magnetic moment also appears naturally in the energy of
the wavepacket when a magnetic field is present.  In the Dirac
Hamiltonian, the magnetic field enters only through the vector
potential, and there is not a Zeeman energy term. By expanding the
vector potential about the center of the wavepacket, we find the
wavepacket energy up to first order in the gradients as
\cite{culcer,ganesh},
\begin{equation}
E(\textbf{r}_c,\textbf{k}_c)=E_0({k}_c)-e\phi(\textbf{r}_c)+\frac{e}{2mc}\textbf{L}
(\textbf{k}_c)\cdot{\textbf B},\label{E}
\end{equation}
where $\textbf{k}_c\equiv\textbf{q}_c+\frac{e}{c}{\textbf
A}(\textbf{r}_c)$, $\textbf{A}$ is the vector potential, and
$E_0({k}_c)=\epsilon({k}_c) mc^2$. The Zeeman energy in
Eq.~(\ref{E}) is just the gradient correction in the wavepacket
energy. It is strikingly observed that there does not exist a
spin-orbit coupling in the wavepacket energy, which one would
normally think to be present to first order in the electric field.
It turns out that the semiclassical dynamics of the Dirac electron
is non-canonical, and spin-orbit coupling is deeply rooted in the
Berry-curvature field which specifies this non-canonical
structure. By requiring the wavepacket to satisfy the
time-dependent Dirac equation, one can derive the effective
Lagrangian \cite{culcer},
\begin{equation}
L_{eff}=i\hbar\eta^\dagger\frac{\partial \eta}{\partial
t}+\hbar\dot\textbf{k}_c\cdot\textbf{R}+\hbar
\textbf{k}_c\cdot{\dot{\textbf
r}_c}-\frac{e}{c}\textbf{A}\cdot{\dot{\textbf{r}}_c}-E(\textbf{r}_c,\textbf{k}_c),
\label{Leff}
\end{equation}
where $\textbf{R}$ is the Berry connection based on the gauge
specified below Eq.~(1), $
\textbf{R}=\frac{\lambda_c^2}{2\epsilon(\epsilon+1)}\textbf{
k}_c\times\langle{\mbox{\boldmath$\sigma$}}\rangle,\label{re2}$ in
which
$\langle\mbox{\boldmath$\sigma$}\rangle\equiv\eta^\dagger\mbox{\boldmath$\sigma$}\eta$.
The equations of motion for the center of charge position and
momentum, correct to linear order in fields, then follow as
\begin{eqnarray}
\hbar\dot\textbf{k}_c&=&-e\textbf{E}-\frac{e}{c}\frac{\hbar\textbf{
k}_c}{\epsilon m}\times\textbf{B},\label{k}\\
\dot\textbf{r}_c&=&\frac{\hbar\textbf{k}_c}{\epsilon
m}+\frac{e}{\hbar}\left(\textbf{E}\times\textbf{F} +\textbf{
B}\cdot\textbf{F}\frac{{\hbar\textbf{k}_c}}{\epsilon
mc}\right),\label{r}
\end{eqnarray}
the Berry curvature $
{\textbf{F}}=-\frac{\lambda_c^2}{2\epsilon^3}\left(\langle\mbox{\boldmath$\sigma$}\rangle+\lambda_c^2\frac{\textbf{
k}_c\cdot\langle\mbox{\boldmath$\sigma$}\rangle}{\epsilon+1}\textbf{k}_c\right)$\label{F}.
These equations need to be solved in conjunction with the equation
for spin precession,
\begin{equation}
\langle\dot{\mbox{\boldmath$\sigma$}}\rangle=\frac{e}{\epsilon mc
}\left[\textbf{B} +\frac{\textbf{E}\times\hbar\textbf{
k}_c}{(\epsilon+1)mc
}\right]\times\langle\mbox{\boldmath$\sigma$}\rangle,\label{S}
\end{equation}
which agrees with the BMT equation\cite{bmt}. The equations of
motion, Eqs.~(\ref{k}),(\ref{r}), and ({\ref S}), are all
invariant under SU(2) gauge transformation.

The same equations of motion have been obtained from a formal
semiclassical expansion to first order in $\hbar$\cite{bliokh},
which is not surprising because our dimensionless weak-field
parameters are $eE\lambda_c/mc^2$ and $eB\lambda_c/mc^2$, which
are both proportional to the Planck constant. Our results are not
limited to low momenta and become even more accurate for high
momenta where the energy gap is greater. For example, one can
obtain the semiclassical but relativistic cyclotron frequency
$\omega_c$ when the electron is confined in a plane perpendicular
to the magnetic field (say $\textbf{E}=0$),
$\omega_c=eBc/E({k}_c)$, where $E({k }_c)=E_0(k_c)+e/(2mc)\textbf{
L}\cdot\textbf{B}$ is the total energy of the electron.

It is evident from the equations of motion that the position and
momentum of the wavepacket do not form a canonical pair, due to
the presence of the gauge potentials $\textbf{R}$ and
$\textbf{A}$. Quantization of the non-canonical equations of
motion for the semiclassical Dirac electron presents an
interesting and important physics problem\cite{bliokh,xiao}. If
one can find new variables $\textbf{r}$ and $\textbf{p}$, such
that the effective Lagrangian is of the following form,
\begin{equation}
L_{eff}=i\hbar\eta^\dagger\frac{\partial\eta}{\partial t}+\textbf{
p}\cdot{\dot\textbf{r}}-E(\textbf{r},\textbf{p}),
\end{equation}
then $\textbf{r}$ and $\textbf{p}$ would naturally be a classical
canonical pair. To linear order of the fields (and up to total
time derivatives), this is indeed possible by the transformation,
\begin{eqnarray}
\textbf{r}&=&\textbf{r}_c-\textbf{R}(\textbf{k}_c)-\textbf{G}(\textbf{k}_c);
\nonumber \\
\textbf{p}&=&\hbar\textbf{k}_c-\frac{e}{c}\textbf{A}(\textbf{r}_c)
-\frac{e}{2c}\textbf{B}\times\textbf{R}(\textbf{k}_c),
\end{eqnarray}
where $G_\alpha\equiv 1/2(\partial \textbf{R}/\partial
k^\alpha)\cdot(\textbf{R}\times\textbf{B})$. Conversely, one can
write
\begin{eqnarray}
\textbf{r}_c&=&\textbf{r}+\textbf{R}(\mbox{\boldmath$\pi$})+\textbf{G}(\mbox{\boldmath$\pi$}),\nonumber\\
\hbar\textbf{k}_c&=&\mbox{\boldmath$\pi$}+\frac{e}{c}\textbf{B}
\times\textbf{R}(\mbox{\boldmath$\pi$}),\label{extra}
\end{eqnarray}
where
$\mbox{\boldmath$\pi$}=\textbf{p}+\frac{e}{c}\textbf{A}(\textbf{r})$.
This is entirely analogous to the Peierls substitution for the
momentum variable. Naively, one expects the familiar presence of
the gauge potentials, $\textbf{R}$ and $\textbf{A}$, to the
position and momentum. However, this is not enough in this
semiclassical formulation and extra corrections are required. The
$\textbf{G}$-term would further shift the position operator, but
does not influence the velocity (to linear order in fields);
similarly, the $\textbf{B}\times\textbf{R}$-term would shift the
momentum, but does not alter the force.

One can {\it re-quantize} the semiclassical Dirac energy by
promoting $\textbf{r}$ and $\textbf{p}$ in Eq~(\ref{E}) to be
quantum conjugate variables, and
$\langle\mbox{\boldmath$\sigma$}\rangle$ in the same equation to
be Pauli matrices. The result turns out to be precisely the
(relativistic) Pauli Hamiltonian accurate to all orders of the
velocity ($\mu_B$ is the Bohr magneton)\cite{silenko},
\begin{equation}
{H}(\textbf{r},\textbf{p})=\epsilon({\pi})
mc^2-e\phi(\textbf{r})+\frac{\mu_B}{\epsilon(\epsilon+1)mc}\mbox{\boldmath$\pi$}\times
\mbox{\boldmath$\sigma$}\cdot\textbf{E}+\frac{\mu_B}{\epsilon}\mbox{\boldmath$\sigma$}
\cdot\textbf{B}\label{pauli}
\end{equation}
That is, one can obtain an effective quantum Hamiltonian by
obtaining the semiclassical energy first, followed by using the
(generalized) Peierls substituion for re-quantization. We
emphasize that one would fail to reproduce the correct Pauli
Hamiltonian if the extra corrections in Eq.~(\ref{extra}) were not
included. This alternative approach is simpler and more intuitive
when compared to formal procedures of block-diagonalization, such
as the Foldy-Wouthuysen transformation.

In Eq.~(\ref{pauli}), the spin-orbit coupling emerges from the
first-order gradient expansion of the scalar potential,
$\partial\phi/\partial\textbf{r}\cdot\textbf{R}$. This reveals a
deep connection between the spin-orbit interaction and the
non-canonical structure of the semiclassical Dirac theory. It must
be remembered that the position variable $\textbf{r}$ in the Pauli
Hamiltonian does not correspond to the true position
$\textbf{r}_c$ of the Dirac electron. The former is sometimes
called the mean position, but it is not really the mean position
of the wave packet. In the literature, $-e\textbf{R}$ has often
been called an electric dipole\cite{yafet}, which couples to the
electric field to give rise to the spin-orbit energy\cite{mathur}.
This is unfortunately artificial, because its existence depends on
an unphysical assignment of the electron position, which depends
on the choice of the SU(2) gauge. Indeed, the equations of motion
based on the Pauli Hamiltonian is consistent with the Dirac theory
if and only if one recognizes this fact.

If the Dirac particle has an anomalous magnetic moment (AMM),
which may be originated from the coupling with quantized
electromagnetic field, or from the internal motion of the
constituent quarks in a nucleon, then in addition to the energy in
Eq.~(\ref{E}), the semiclassical particle acquires an extra energy
($a\equiv g/2-1$),
\begin{eqnarray}
E_{\rm AMM}&=&a\mu_B\langle
w|\beta(-i\mbox{\boldmath$\alpha$}\cdot\textbf{E}
+\mbox{\boldmath$\Sigma$}\cdot\textbf{B})|w\rangle\nonumber\\
&=&a\mu_B\left[\frac{\textbf{k}_c\times\langle\mbox{\boldmath$\sigma$}
\rangle\cdot\textbf{E}}{\lambda_c^{-1}\epsilon}+\left(\langle
\mbox{\boldmath$\sigma$}\rangle-\frac{(\textbf{k}_c
\cdot\langle\mbox{\boldmath$\sigma$}\rangle)\textbf{k}_c}{\lambda_c^{-2}
\epsilon(\epsilon+1)}\right)\cdot\textbf{B}\right]\label{AMM}\cr
\end{eqnarray}
Unlike the energy in Eq.~(\ref{E}), the spin-orbit coupling is now
explicit from the very beginning. It indicates that, when the
particle has a AMM, the true center of charge $\textbf{r}'_c$ has
been further displaced from $\textbf{r}_c$. In order to obtain the
spin-orbit interaction in Eq.~(\ref{AMM}), the amount of
displacement has to be $\textbf{
R}'=a({\lambda_c^2}/{2\epsilon})\textbf{
k}_c\times\langle\mbox{\boldmath$\sigma$}\rangle$. Indeed, if one
replaces the $\textbf{r}_c$ and $e/(2mc)$ in Eq.~(\ref{E}) with
$\textbf{r}'_c=\textbf{r}_c+\textbf{R}'$ and $eg/(4mc)$, then the
AMM energy of the wavepacket can be fully reproduced from the
simpler semiclassical energy in Eq.~(\ref{E}). Furthermore,
following the same re-quantization scheme using the generalized
Peierls substitution, one can indeed obtain the (relativistic)
Pauli Hamiltonian with the exact AMM terms (Eq.~(33) in
Ref.~\cite{silenko}).

We emphasize that, since one expects no electric dipole from a
point particle, the presence of the electric dipole energy
indicates that the theory is only an effective one, and the
position of the particle in the effective theory should differ
from the true position by a Berry connection. Of course, in
predicting the trajectory of the point particle, one always has to
refer to its true position. Therefore, in the Dirac theory with
$H_{\rm AMM}=a\mu_B\beta(-i\mbox{\boldmath$\alpha$}\cdot\textbf{E}
+\mbox{\boldmath$\Sigma$}\cdot\textbf{B})$, the true position
$\textbf{r}$ is conjectured to be $\textbf{r}_D+\textbf{R}_D$,
where $\textbf{r}_D$ is the usual position operator, and
$\textbf{R}_D=(a\mu_B/e)(-i\beta\mbox{\boldmath$\alpha$})$ is the
displacement required to generate the electric field term in
$H_{\rm AMM}$. The displacement (Berry connection) is originated
from the projection of a larger state space (for example, the one
that includes the quantized electromagnetic field) to the Hilbert
space of the relativistic particle. Thus, the true velocity
operator of the particle should be,
\begin{equation}
{\dot\textbf{r}_D}+{\dot\textbf{R}_D}
=(1+a)c\mbox{\boldmath$\alpha$}-a\beta\frac{\mbox{\boldmath$\pi$}}{m}
+a^2\frac{\mu_B}{mc}\left(\textbf{E}\times\mbox{\boldmath$\Sigma$}
+\gamma_5\textbf{B}\right).
\end{equation}
The magnitude of the anomaly $a$ for an electron is of the order
of $10^{-3}$ and might be too small for such a difference to be
observed. However, the anomaly for a nucleon is of order one.
Therefore, it is possible to observe and verify such a deviation
in experiments using relativistic proton beams.\cite{a}

In summary, we demonstrate an alternative method to investigate
the Dirac electron by semiclassical wavepacket approach which
provides a much more intuitive and pedagogic picture of the Dirac
electron. In this framework, the intrinsic size of the
non-relativistic electron is of the order of $\lambda_c$, and the
self-rotation motion generates a magnetic moment equals precisely
the conventional spin magnetic moment. The dynamics of the
wavepacket is also studied and reveals a direct link between the
Berry connection and the spin-orbit coupling. By re-quantizing the
semiclassical Dirac theory through a generalized Peierls
substitution, we recover the relativistic Pauli Hamiltonian to all
orders of velocity.

The authors would like to thank E. I. Rashba, M. Stone, L. Balent,
D. Culcer, D. Xiao, and M.F. Yang for many helpful discussions.


\begin{thebibliography}{99}
\bibitem{a}
However, when the velocity is projected to the subspace for an
electron with positive energy, one has $P
c\mbox{\boldmath$\alpha$}P=\mbox{\boldmath$\pi$}/(\epsilon m)$. It
would cancel with the term $P
\beta\mbox{\boldmath$\pi$}/mP=\mbox{\boldmath$\pi$}/m$ to leading
order of the velocity. Therefore, the leading (field-free)
correction is of the order of $a(v/c)^2$.
\end{thebibliography}
\end{document}